# Modeling the Oblique Spin Precession in Lateral Spin Valves for Accurate Determination of Spin Lifetime Anisotropy: Effect of Finite Contact Resistance and Channel Length

Tiancong Zhu and Roland K. Kawakami

*Department of Physics, The Ohio State University, Columbus, OH  43210, USA*

**Abstract**

The spin lifetime anisotropy is an important quantity for investigating the spin relaxation mechanisms in graphene and in heterostructures of two-dimensional materials. We generalize the diffusive spin transport equations of oblique spin precession in a lateral spin valve with finite contact resistance. This yields a method to determine the spin lifetime anisotropy ratio $\xi = \tau_\perp/\tau_\parallel$, which is the ratio between lifetimes of spin polarized perpendicular and parallel to the graphene surface. By solving the steady-state Bloch equations, we show that the line-shape of the oblique spin precession signal can be described with six dimensionless parameters, which can be solved analytically. We demonstrate that the anisotropic spin precession characteristics can be strongly suppressed by contact induced spin relaxation originating from conductance mismatch between the channel material and electrodes. To extract the spin lifetime anisotropy ratio accurately, we develop a closed form equation that includes the effect of finite contact resistance. Furthermore, we demonstrate that in the high contact resistance regime, the minimum channel length required for accurately determining the spin lifetime anisotropy for a sufficiently low external magnetic field is only determined by the diffusion coefficient of the channel material, as opposed to the spin diffusion length. Our work provides an accurate model to extract the spin lifetime anisotropy ratio from the oblique spin precession measurement, and can be used to guide the device design for such measurements.

E-mail: zhu.1073@osu.edu, kawakami.15@osu.edu

# I. INTRODUCTION

Spintronics aims to utilize the spin degree freedom of charge carriers for logic operation and information storage [1]. In recent years, graphene has emerged as one of the most efficient spin channel materials [2], exhibiting gate tunable spin transport, long spin lifetimes and long spin diffusion lengths at room temperature [3-12]. These make graphene a promising material for spintronics applications [13-19]. What makes graphene even more special is the high tunability of its properties. Due to the atomically thin nature of graphene, its properties are strongly subject to the environment, such as surface flatness [20-25], adatom adsorption [26-33], or in proximity with other materials [34-50]. This allows manipulation of graphene's spin transport and magnetic properties, which further enriches the possibilities of graphene for spintronics.

Among all the properties in graphene, spin-orbit coupling is of particular interest. The intrinsic spin-orbit coupling in graphene is predicted to be very weak, with a magnitude of only $\sim 30~\mu eV$ [51-53]. However, this value can be enhanced by several orders of magnitude by modifying graphene surface with adatoms, hybridizing with metal, or in proximity with strong spin-orbit coupling material [35,37-40,47,54-57]. Such strong spin-orbit coupling interaction is essential for new phenomena, such as spin Hall effect (SHE) [58-63], anomalous Hall effect (AHE) [36,64], quantum spin Hall effect (QSHE) [65,66] and quantum anomalous Hall effect (QAHE) [67-71] to appear in graphene. Some of the above effects have been observed in experiments [35,36,60,61,64]. Furthermore, spin-orbit coupling can play a crucial role in the spin relaxation mechanism in graphene [72-76]. Up to now, the experimentally observed spin lifetime (12 ns, in [7]) in graphene is still two orders of magnitude smaller than the theoretical predictions (~1 μs, [2]). While the dominating spin relaxation mechanism in graphene remains unclear, spin relaxation through spin-orbit coupling is one major candidate. A careful study of spin-orbit coupling will be beneficial for determining the key limiting factors of spin transport in graphene.

One consequence that spin-orbit coupling produces in graphene is the spin lifetime anisotropy, in which case the spin polarization perpendicular and parallel to the graphene sheet have different lifetimes [77]. Conventionally, the spin lifetime anisotropy ratio, $\xi = \tau_\perp/\tau_\parallel$ is used to describe this phenomenon. The spin lifetime anisotropy originates from spin relaxation dominated by a directional spin orbit field (SOF): For the Rashba type of SOF, which lies parallel to the graphene sheet, $\xi < 1$ is expected; for the Kane-Mele type of SOF, which is perpendicular to the graphene sheet, $\xi > 1$ is expected. In the case for other spin relaxation mechanisms, such as resonant scattering from magnetic impurities, an isotropic spin relaxation is expected. Observing an anisotropic spin relaxation in graphene is the fingerprint of spin-orbit driven spin relaxation [78].

Spin lifetime anisotropy was originally measured in graphene by applying a large magnetic field perpendicular to the graphene surface [79,80]. The applied magnetic field magnetizes the ferromagnetic electrodes into the field direction, which allows out-of-plane spin injection. However, it typically requires $> 1\,T$ of magnetic field to fully magnetize the electrodes out-of-plane. Such a large magnetic field can cause side effects, such as ordinary magneto-resistance, that may contribute significantly to the signal. Recently, Raes *et al.* have demonstrated a new method to measure spin lifetime anisotropy in graphene with oblique spin precession in the lateral spin valve geometry [78]. In this geometry, an oblique magnetic field with relatively small magnitude (typically ~150 mT) is applied, and spin precession signal is measured. The oblique magnetic field makes spin in the graphene channel precess into the out-of-plane direction, thus sampling both the in-plane and out-of-plane components of spin relaxation. The much smaller magnitude of applied magnetic field avoids side effects mentioned previously, which allows a more accurate measurement on the anisotropy ratio. However, two important issues still need to be addressed for the oblique spin precession measurement. First, the finite contact resistance between the ferromagnetic electrodes and graphene in the lateral spin valve can act as a spin sink, which has been shown to cause an underestimation of spin lifetime extracted from spin precession measurement [81-84]. Such underestimation can also exist in oblique spin precession measurement. A quantitative method should be introduced to account for such an effect. Second, to perform oblique spin precession measurement with small magnetic fields, it has been assumed that a relatively long spin diffusion channel is required, previously estimated as $L \geq \sqrt{2}\lambda_s = \sqrt{2D\tau_s}$ [78]. However, this makes the oblique spin precession method seemingly unsuitable for graphene devices with long spin lifetimes due to the requirement of extremely long device channels. To our knowledge, neither of these two issues have been thoroughly discussed.

In this paper, we present our model on oblique spin precession in the lateral spin valve geometry to address the above two issues. First, we develop an analytical method for calculating the spin precession curves with finite contact resistance, and obtain a closed form expression for extracting the spin lifetime anisotropy ratio from the measurement. This provides a method for accurately determining the spin lifetime anisotropy in realistic lateral spin valve devices with finite contact resistance. Furthermore, we derive a closed form expression to determine the minimum channel length required for oblique spin precession measurement. Our result shows that only a moderate length of the spin channel is needed for graphene and is determined by the diffusion coefficient as opposed to the spin diffusion length. Overall, our result provides a means to extract the spin lifetime anisotropy ratio from the oblique spin precession geometry and also serves as a guide for designing devices for such a measurement. This formalism can also be applied to other channel materials such as graphene-transition metal dichalcogenide heterostructures, which have recently exhibited strong spin lifetime anisotropy [49,50].

## II. MODELING DETAILS

The oblique spin precession measurement is performed in the non-local geometry. Figure 1(a) shows the schematics of such a device. To achieve spin transport, an electric current is applied from the left ferromagnetic (FM) electrode (injector) into the channel, which builds up spin accumulation underneath the injector. The spin accumulation can diffuse across the channel and reach the right FM electrode (detector). Depending on the magnitude and polarization direction of the diffusive spin accumulation relative to the FM detector electrode, a high (low) voltage signal can be measured at the detector. This voltage signal is the so-called non-local voltage ($V_{NL}$), resulting from spin transport in the channel material.

To perform oblique spin precession measurement, an external magnetic field is applied in the y-z plane, with an angle $\beta$ from the channel surface (Figure 1(b)). The spin in the material precesses around the magnetic field while diffusing through the channel. The precession results in a reduction of $V_{NL}$ as a function of applied field. A plot of $V_{NL}$ as a function of magnetic field is defined as the non-local spin precession curve. For a material with anisotropic spin relaxation, the line-shape of the non-local spin precession curve will depend on the applied field angle, which is a signature of spin lifetime anisotropy. Furthermore, when the magnetic field is large enough ($B_{Sat}$), the spin polarization perpendicular to the field will be fully dephased, and the signal will be saturated with the component parallel to the field. The curvature of $V_{NL}$ as function of field angle β can be used to determine the value of ξ. Both the line-shape of the non-local spin precession at different oblique angle and curvature of signal in the saturation limit are important for identifying the spin lifetime anisotropy in the spin diffusion channel.

To model oblique spin precession in lateral spin valves, we employ the one-dimensional steady state Bloch equation to describe spin transport in the device channel

$$D\nabla^2 \vec{\mu}^s - \gamma_c \cdot \vec{\mu}^s \times \vec{B} - \overline{\tau_s^{-1}} \cdot \vec{\mu}^s = 0 \tag{1}$$

where $D$ is the spin diffusion coefficient of the channel, $\gamma_c$ is the gyro-magnetic ratio of the charge carrier, and $\vec{B}$ is the oblique magnetic field. The spin dependent chemical potential $\vec{\mu}^s$ is a three-component vector, with each of the component describing the spin population projected along the corresponding Cartesian axes. The spin relaxation matrix $\overline{\tau_s^{-1}}$ describes the spin lifetime anisotropy with different in-plane and out-of-plane spin relaxation rates.

A natural way to solve Eq. (1) is to transform to the Cartesian frame $(e_x, e_{B_\parallel}, e_{B_\perp})$ that is affixed with the applied field (figure 1b). This is because the applied magnetic field can only induce precession to the spin population perpendicular to it. In the new frame, $\vec{B} = (0, B, 0)$, $\vec{\mu}^s = (\mu_x^s, \mu_{B_\parallel}^s, \mu_{B_\perp}^s)$, and the spin relaxation matrix $\overline{\tau_s^{-1}}$ can be written as

$$\overline{\tau_s^{-1}} = \tau_\parallel^{-1} \begin{bmatrix} 1 & 0 & 0 \\ 0 & 1 + f(\xi)\sin^2(\beta) & f(\xi)\sin(\beta)\cos(\beta) \\ 0 & f(\xi)\sin(\beta)\cos(\beta) & 1 + f(\xi)\cos^2(\beta) \end{bmatrix} \tag{2}$$

with $f(\xi) = 1/\xi - 1$.

We find that the steady-state Bloch equation in the new Cartesian frame can be solved analytically. By performing the Fourier transform to Eq. (1), we obtain

$$\begin{bmatrix} \lambda_\parallel^2 k^2 + 1 & 0 & -\tau_\parallel \gamma_c B \\ 0 & \lambda_\parallel^2 k^2 + 1 + f(\xi)\sin^2(\beta) & f(\xi)\sin(\beta)\cos(\beta) \\ \tau_\parallel \gamma_c B & f(\xi)\sin(\beta)\cos(\beta) & \lambda_\parallel^2 k^2 + 1 + f(\xi)\cos^2(\beta) \end{bmatrix} \begin{bmatrix} \mu_x^s \\ \mu_\parallel^s \\ \mu_\perp^s \end{bmatrix} = 0 \tag{3}$$

where $\lambda_\parallel = \sqrt{D\tau_\parallel}$ is the spin diffusion length for spin polarized in-plane. Solving Eqn. (3) leads to the general solution of the spin dependent chemical potential

$$\mu_\nu^s = \sum_n C_{n,\nu}^\pm \exp(-ik_n^\pm x) \tag{4}$$

where $n = 1, 2, 3$ numbers the three modes (defined below), $k_n^{\pm}$ are the corresponding wave vectors, and $v = e_x, e_{B_{\parallel}}, e_{B_{\perp}}$ represent the different spatial components of the spin dependent chemical potential. To obtain the expression for Eq. (3), we define $K = \lambda_{\parallel}^2 k^2 + 1$ and solve Eq. (3). This leads to three different non-zero modes of spin diffusion in the channel given by

$$K_n = \alpha + e^{i\frac{n\pi}{3}}\left(\beta + (\beta^2 + \gamma^3)^{\frac{1}{2}}\right)^{\frac{1}{3}} + e^{-i\frac{n\pi}{3}}\left(\beta - (\beta^2 + \gamma^3)^{\frac{1}{2}}\right)^{\frac{1}{3}} \quad n = 1,2,3 \quad (5)$$

and

$$\begin{cases} \alpha = -\frac{f(\xi)}{3} \\ \beta = \frac{b^2 f(\xi)}{6} - \frac{f^3(\xi)}{27} - \frac{b^2 f(\xi) \sin^2(\beta)}{2} \\ \gamma = \frac{b^2}{3} - \frac{f^2(\xi)}{9} \end{cases} \quad (6)$$

Each mode in Eq. (5) contains two wave vectors ($k_n^{\pm} = \pm\lambda_{\parallel}^{-1}\sqrt{K_n - 1}$, with $n = 1,2,3$). The $k_n^{+(-)}$ are complex numbers with a positive (negative) imaginary part, corresponding to a wave that decays [imaginary part] while oscillating [real part] as it transports to the $-(+)x$ direction. The general solution in Eq. (4) can then be written as

$$\mu_v^s = \sum_{n=1,2,3}\left(c_n^+ \Gamma_v(K_n) e^{-i\lambda_{\parallel}^{-1}\sqrt{K_n-1}\cdot x} + c_n^- \Gamma_v(K_n) e^{+i\lambda_{\parallel}^{-1}\sqrt{K_n-1}\cdot x}\right) \quad (7)$$

with

$$\begin{cases} \Gamma_x(K_n) = -b[K_n + f(\xi)\sin^2(\beta)] \\ \Gamma_{B_{\parallel}}(K_n) = K_n f(\xi) \cos(\beta) \sin(\beta) \\ \Gamma_{B_{\perp}}(K_n) = -K_n[K_n + f(\xi)\sin^2(\beta)] \end{cases} \quad (8)$$

Eq. (7) fully describes the spin accumulation in each region of the spin transport channel between the ferromagnetic electrodes.

To include the effect of spin absorption at both of the FM electrodes, we consider the continuity equation in the spin diffusion channel underneath the electrodes. For each FM electrode,

$$\begin{cases} \mu_\nu^s(x_c^-) = \mu_\nu^s(x_c^+) \\ I_\nu^s(x_c^-) = I_{\nu(\text{abs.})}^s(x_c) + I_\nu^s(x_c^+) \end{cases} \quad (9)$$

The first equation relates to continuity of the spin dependent chemical potential, and the second equation represents the continuity of the spin current. The $x_c$ in the equation is the position of the FM contact, with the superscript $+(-)$ represents the position of the channel just to the right(left) of the contact. Assuming the spin absorption current into the FM, $I_{\nu(\text{abs.})}^s(x_c)$, is isotropic with the spin polarization, both the spin current in the channel material as well as the absorption current can be expressed with $\mu_\nu^s(x_c)$ underneath the channel as

$$\begin{cases} I_\nu^s = -W \cdot \sigma \cdot \nabla \mu_\nu^s(x) \\ I_{\nu(\text{abs.})}^s(x_c) = \tilde{R}_c^{-1} \cdot \mu_\nu^s(x_c) + P \cdot I^c \end{cases} \quad (10)$$

where W is the width of the channel, $\sigma$ is the electrical conductivity of the channel, and $I^c$ is the charge current flow through the FM electrode. $\tilde{R}_c = R_c/(1-P^2) + R_F/(1-P^2)$ is the effective contact resistance of the electrode, with $R_c$ the interfacial resistance, $R_F = \lambda_F \rho_F / A_F$ the spin resistance of the electrode, and $P$ the spin polarization of the FM electrode. Combining Eqs. (7), (9) and (10) generates a fully defined system of linear equations, and $\mu_\nu^s(x)$ can be solved analytically at any given position and external magnetic field. The non-local voltage can then be extracted from $\mu_\nu^s(x)$

$$V_{NL} = P_{\text{det}} \cdot \left(\mu_{B_\parallel}^s(x_{det}) \cdot \cos(\beta) - \mu_{B_\perp}^s(x_{det}) \cdot \sin(\beta)\right) \quad (11)$$

To describe the line-shape of the oblique spin precession, we normalize the non-local signal with its zero-field value

$$V_{NL}^* = \frac{V_{NL}(B)}{V_{NL}(B=0)} \quad (12)$$

we find that $V_{NL}^*$ can be fully described by six dimensionless parameters: magnetic field angle $\beta$, anisotropy ratio $\xi$, normalized channel length $l = L/\lambda_\parallel$, normalized magnetic field strength $b = B/(\tau_\parallel \gamma_c)^{-1}$, normalized contact resistance for injector $r_{inj} = \tilde{R}_{inj}/R^s$, and detector $r_{det} = \tilde{R}_{det}/R^s$. Here $L$ is the channel length between the injector and detector, $\lambda_\parallel$ is the diffusion length for spin polarized in-plane, and $R^s = R_{Gr} \cdot \lambda_\parallel/L$ is the spin resistance of the channel. This provides a generic description for oblique spin precession in materials with spin lifetime anisotropy.

## III. RESULTS AND DISCUSSION

### A. Anisotropic spin precession with finite contact resistance

We first discuss the effect of finite contact resistance on the oblique spin precession signal with spin lifetime anisotropy. Figure 2 shows a set of non-local spin precession curves generated with different anisotropy ratio $\xi$ and normalized contact resistance $r = r_{inj} = r_{det}$. The external magnetic field is set to be $\beta = 45°$ from the sample surface. In the case of large contact resistance (Figure 2(a)), the effect of spin absorption is suppressed, and a significant variation in the line-shape of the non-local spin signal as function of spin lifetime anisotropy ratio $\xi$ is observed. As the contact resistance decreases, the conductance mismatch between the electrode and channel becomes prominent, and the effect of anisotropic spin precession is suppressed. When the device enters into the transparent contact regime, which is shown in figure 2(b), the contact induced spin relaxation dominates the overall spin relaxation. Under this condition, a set of wider spin precession curves is observed, with much less variation at different spin lifetime anisotropy ratio.

The effect of finite contact resistance on anisotropic spin precession can be seen more clearly with the angular dependence of $V_{NL}^*(b \to \infty)$. By taking the approximation that $b \to \infty$, a closed form solution of $V_{NL}^*$ can be calculated from our model

$$V_{NL}^*(b \to \infty) = \frac{g(\beta,\xi) \cdot [(1+2r_{inj})(1+2r_{det})-e^{-2l}]}{(1+2g(\beta,\xi) \cdot r_{inj})(1+2g(\beta,\xi) \cdot r_{det})-e^{-2l \cdot g(\beta,\xi)}} \cdot e^{-l \cdot (g(\beta,\xi)-1)} \cdot \cos^2(\beta) \qquad (13)$$

with $g(\beta,\xi) = \left(1-\frac{1}{\xi}\right)\cos^2(\beta) + \frac{1}{\xi}$. Eq. (13) should be used to fit the oblique Hanle signal at saturation vs. $\cos^2(\beta)$ to extract an accurate value for the spin lifetime anisotropy ratio $\xi$. To check that this equation encompasses the previous model that does not include spin absorption, we take the limit of high contact resistance ($r_{inj}, r_{det} \gg 1$), and the expression can be simplified as

$$V_{NL}^*(b \to \infty) = g^{-1}(\beta,\xi) \cdot e^{-l \cdot (g(\beta,\xi)-1)} \cdot \cos^2(\beta) \qquad (14)$$

Eq. (14) is the same as that in [78,85].

Figure 3 shows the angle dependent $V_{NL}^*(b \to \infty)$ curves with different channel length and contact resistance generated with Eq. (11). In the case of a moderate channel length ($L = \lambda_\parallel$), the effect of anisotropic spin precession is clear in a high contact resistance device (figure 3(a)). However, the curves with different anisotropy ratio almost collapse onto a straight line representing $\xi = 1$ as the device enters the transparent contact regime (figure 3(b)). With a much longer channel length ($L = 3\lambda_\parallel$, figure 3(c, d)), the effect of anisotropic spin precession becomes more prominent, and the suppression of anisotropic spin precession due to the low contact resistance becomes less effective. However, there is still an obvious discrepancy between high contact resistance and transparent contact devices. In terms of experiment, such a discrepancy from contact induced spin relaxation can lead to a strong underestimation of the spin lifetime anisotropy ratio in the oblique spin precession measurement.

The suppression of anisotropic spin precession can be understood by contact induced spin relaxation, which originates from the conductance mismatch between the electrode and spin diffusion channel. The spin current absorbed by an electrode with finite contact resistance can be

expressed as

$$I_{abs.}^s = \mu^s/\tilde{R}_c \tag{15}$$

where $\mu^s$ is the spin dependent chemical potential under the electrode, and $\tilde{R}_c$ is the effective contact resistance defined previously. The spin absorption acts as an additional spin sink for the spin population in the channel, which is equivalent to adding another relaxation source. Since we assume isotropic spin absorption, the additional contact induced spin relaxation will tend to bring the measured anisotropy ratio back to unity. For a device with high contact resistance, the contact induced spin relaxation is weak and the oblique spin precession is measuring mainly the intrinsic anisotropy ratio of the channel. However, when the contact is more transparent-like, the isotropic contact induced spin relaxation will dominate the signal, and make the anisotropic spin precession feature less obvious.

To illustrate how contact induced spin relaxation affects extracting the spin lifetime anisotropy ratio from the oblique spin precession measurement, we perform the following simulation. First, we generate a set of angular dependent $V_{NL}^*(b \to \infty)$ curves of different channel anisotropy ratio $\xi_{real}$ and finite contact resistance with Eq. (13), then use Eq. (14) to fit the simulated curves while ignoring spin absorption effects and extract $\xi_{fit}$. The difference between $\xi_{real}$ and $\xi_{fit}$ can be used to quantify the effect of contact induced spin relaxation on oblique spin precession measurement. Figure 4(a) shows one example of the fitting process, where the normalized contact resistance is chosen to be $r = 1$ for generating the simulated curves. Without considering the effect of contact resistance, Eq. (14) can still fit the line-shape of the generated angular dependent curves very well, but the fitted anisotropy ratio $\xi$ is consistently underestimated. In the case that $\xi_{real} = 2.0$, an underestimation of more than 15% is observed.

To further understand the effect of contact induced spin relaxation, we perform the same fitting procedures with a wide range of normalized resistance $r$ and channel length $l$. Figure 4(b) shows the result with $\xi = 2$. The difference of $\Delta\xi = |\xi_{fit} - \xi_{real}|$ normalized with the simulated $\xi_{real}$ is shown on the graph. As seen in the plot, the accuracy of the extracted anisotropic spin precession measurement is mostly dominated by the normalized contact resistance. In the transparent regime, a discrepancy of more than 40% can be observed. The channel length has some influence in reducing $\Delta\xi$, but the overall impact is limited. Our result shows that in order to accurately extract the spin lifetime anisotropy, the contact induced spin relaxation has to be minimized, and a model which considers finite contact resistance is preferred for analyzing the data.

**B. Determining the minimum channel length required for a sufficiently low saturation magnetic field $B_{Sat}$**

In order to saturate the oblique spin precession signal at a sufficiently low external magnetic field $B_{sat}$, the spin diffusion channel must be longer than a minimum channel length. This sets a special requirement in lateral spin valve fabrication for performing such a measurement. However, to date, there is no clear analytical study for the relationship between the minimum required channel length and the corresponding saturation magnetic field. In the following section, we show our approach in understanding this problem.

We first discuss the case that $r \gg 1$, so the contact induced spin relaxation is minimized. This is the ideal case for oblique spin precession measurement, as discussed previously. In order to determine the relationship between $B_{sat}$ and the minimum channel length, we derive the expression of $V_{NL}^{\perp}$, the non-local signal contribution from spin perpendicular to the magnetic field. Combining Eqs. (7), (9) and (10), and assuming that the magnetic field is large enough ($b \gg 1$),

$V_{NL}^{\perp}$ can be written as

$$V_{NL}^{\perp}/I = P_{inj}P_{det}R_s \sin^2(\beta) \cdot \left(\sqrt{2b}\right)^{-1} \cdot \cos\left(-l \cdot \sqrt{b/2} - \pi/4\right) \cdot e^{-l\sqrt{b/2}} \quad (16)$$

Figure 5(a) plots the $V_{NL}$ vs. $b$ curve from Eq. (16) (dashed line) and compares it with the $V_{NL}$ vs. $b$ curve from the general result (i.e. Eq. (11) with $\beta = 90°$) (solid line). The agreement between the two curves for high fields (e.g. $b > 5$) indicates that Eq. (16) describes the high field behavior of the spin precession curve very well. In the saturated limit, the perpendicular component of spin should be fully dephased and the magnitude of $V_{NL}^{\perp}$ should be negligible. We notice that the last exponential term $e^{-l\sqrt{b/2}}$ in Eq. (16) determines the overall magnitude of $V_{NL}^{\perp}$. Similar as Eq. (12), the relative magnitude of $V_{NL}^{\perp}$ compared to the total non-local voltage at zero magnetic field can be written as

$$V_{NL}^{\perp *} \propto e^{-l(\sqrt{b/2}-1)} \quad (17)$$

Defining a threshold value for saturation as $V_{sat}^* = V_{NL}^*(b = 0) \cdot 10^{-\eta}$, the condition for saturation (i.e. $V_{NL}^{\perp *} \leq V_{sat}^*$) generates a requirement for the channel length to be

$$l(\sqrt{b/2} - 1) \geq \eta \ln 10 \quad (18)$$

Considering that $b = B/(\tau_{\parallel}\gamma_c)^{-1} \gg 1$ and $l = L/\lambda_{\parallel} = L/\sqrt{D\tau_{\parallel}}$, one can derive that

$$L \geq \frac{\sqrt{2D}}{\sqrt{B_{sat}\gamma_c}} \cdot \eta \ln 10 = \sqrt{2D(B_{sat}\gamma_c)^{-1}} \cdot \eta \ln 10 \quad (19)$$

We notice that from Eq. (19), the minimum channel length is only determined by the diffusion coefficient and $B_{sat}$. This result can be understood as follows: in the oblique spin precession measurement, the spin relaxation rate is determined by both the intrinsic mechanism and the spin dephasing due to the external magnetic field. In the limit $\tau_{\parallel} \gg (B_{sat}\gamma_c)^{-1}$, spin relaxation due to dephasing dominates, resulting in an effective spin diffusion length $\lambda_{eff} = \sqrt{D\tau_B} \propto \sqrt{D(\gamma_c B_{sat})^{-1}}$. To fully minimize the signal from the perpendicular spin population, $L/\lambda_{eff} \propto \sqrt{\gamma_c L^2 B_{sat}/D} \gg 1$. This leads to $L \gg \sqrt{D(\gamma_c B_{sat})^{-1}}$, which is the same as the result in Eq. (19).

The derivation of Eq. (19) requires the assumption that $b = \tau_\parallel/(B_{sat}\gamma_c)^{-1} \gg 1$. We justify that such an assumption is almost always valid in graphene lateral spin valves. Choosing $B_{sat} = 150\ mT$ as a typical low value to be used in oblique spin precession measurement, we can calculate $(B_{sat}\gamma_c)^{-1} = 38\ ps$, which is indeed much smaller than the spin lifetime of graphene that are currently observed in experiments.

Eq. (19) shows that the oblique spin precession is still a good method to determine spin lifetime anisotropy for graphene with long spin lifetime and spin diffusion length. For example, in a device with currently the highest reported spin lifetime and spin diffusion length ($12\ ns$, $30\ \mu m$ as shown in [7]), the diffusion coefficient can be calculated as $D = \frac{\lambda^2}{\tau} = 0.075\ m^2 s^{-1}$. Assuming that the threshold $10^{-\eta} = 10^{-3}$ and $B_{sat} = 150\ mT$, a minimum channel length of $16.5\ \mu m$ is needed. This channel length is much more feasible for device fabrication and characterization compared to previously estimated $\sqrt{2}\lambda = 42.4\ \mu m$. [7,78]

Finally, we discuss the minimum channel length for a lateral spin valve over a range of contact resistances. We discuss the effect through plotting the spin precession curves $V_{NL}^{\perp\ *}(B)$ with different normalized contact resistances. Figure 5(b) shows a set of such plots, assuming $\tau_\parallel = 12\ ns$, $D = 0.075\ m^2 s^{-1}$, and $L = 16.5\ \mu m$. As shown in the figure, $B_{sat} = 150\ mT$ is enough to saturate the spin signal even with $r = 1$ (green curve), thus the criterion we developed in Eq. (19) is still valid for devices with moderate contact resistance. However, the same $B_{sat}$ is clearly not enough to fully dephase the spin signal when $r$ is even smaller (red and blue curve). This can be understood as following: As the contact resistance decreases, the contact induced spin relaxation starts dominating the spin transport, making the observed spin lifetime $\tau_{ob}$ shorter than $\tau_\parallel$. When the contact induced spin relaxation is strong enough, $\tau_{ob}$ will be greatly suppressed and the assumption $b = \tau_{ob}/(B_{sat}\gamma_c)^{-1} \gg 1$ becomes not valid anymore. As a result, a longer channel length is required for lateral spin valves with strong contact induced spin relaxation. The result is the same for devices with different parameters according to our simulation. This further shows that a non-local spin valve with tunnel barrier is preferred for oblique spin precession measurement.

## IV. CONCLUSIONS

We have proposed a model based on steady-state Bloch equation to describe oblique spin precession in lateral spin valve. Our model considers the effect of finite contact resistance on spin lifetime anisotropy measurement. We demonstrate that the contact induced spin relaxation can strongly suppress the anisotropic spin precession signature in the measurement, which can lead to underestimation of the spin lifetime anisotropy. To solve this issue, we develop a closed form equation for extracting the spin lifetime anisotropy ratio, which accounts for the effect of finite contact resistance. Furthermore, we also derived the relationship between saturation magnetic field $B_{sat}$ and the minimum required channel length. We show that for graphene lateral spin valves, the minimum required channel length is mostly determined by both $B_{sat}$ and the diffusion coefficient of the channel. As a result, the oblique spin precession measurement is suitable for studying graphene lateral spin valves with long spin lifetimes.

**ACKNOWLEDGEMENTS**

We acknowledge the technical assistance of Dongying Wang and support from the US Department of Energy (Grant No. DE-SC0018172).

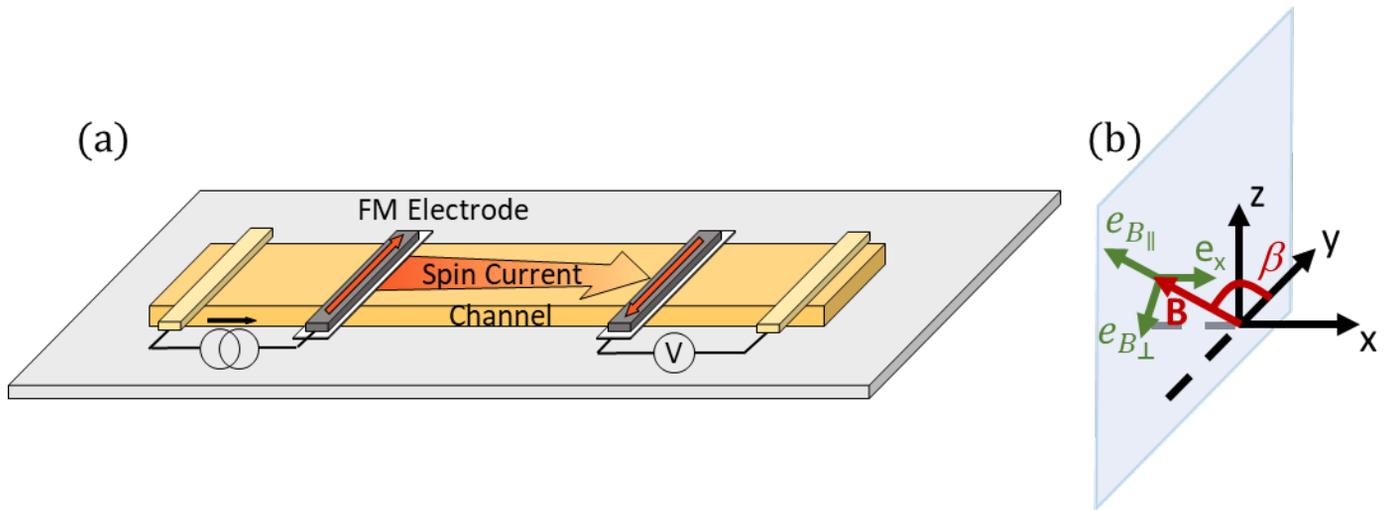

Fig. 1. (a) Schematic drawing of non-local spin valve geometry under magnetic field. For conventional Hanle measurement, the magnetic field is perpendicular to the channel material ($\beta = 90°$). In the oblique spin precession measurement, $\beta$ is varied between 0°~90°, and a magnetic field dependent non-local voltage is measured. (b) The Cartesian frame used in the modeling.

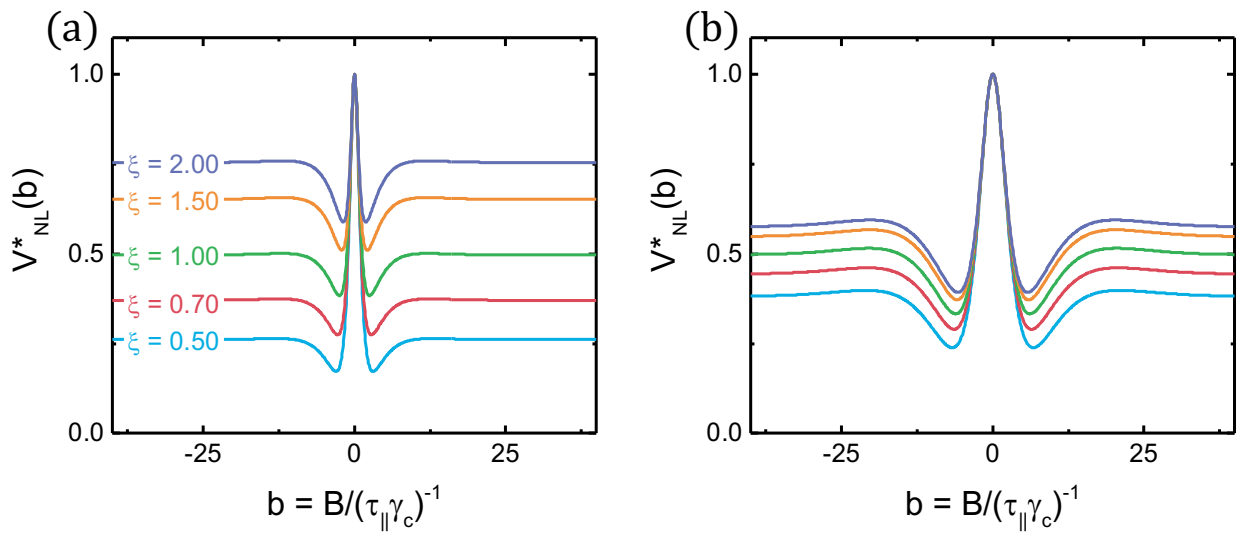

Fig. 2. (a, b) Non-local spin precession curves with high contact resistance ($r = 100$) and transparent contacts ($r = 0.01$), respectively. Curves in each figures from top to bottom correspond to $\xi = 2.0, 1.5, 1.0, 0.7, 0.5$. All the curves are generated with parameters $l = 2$ and $\beta = 45°$.

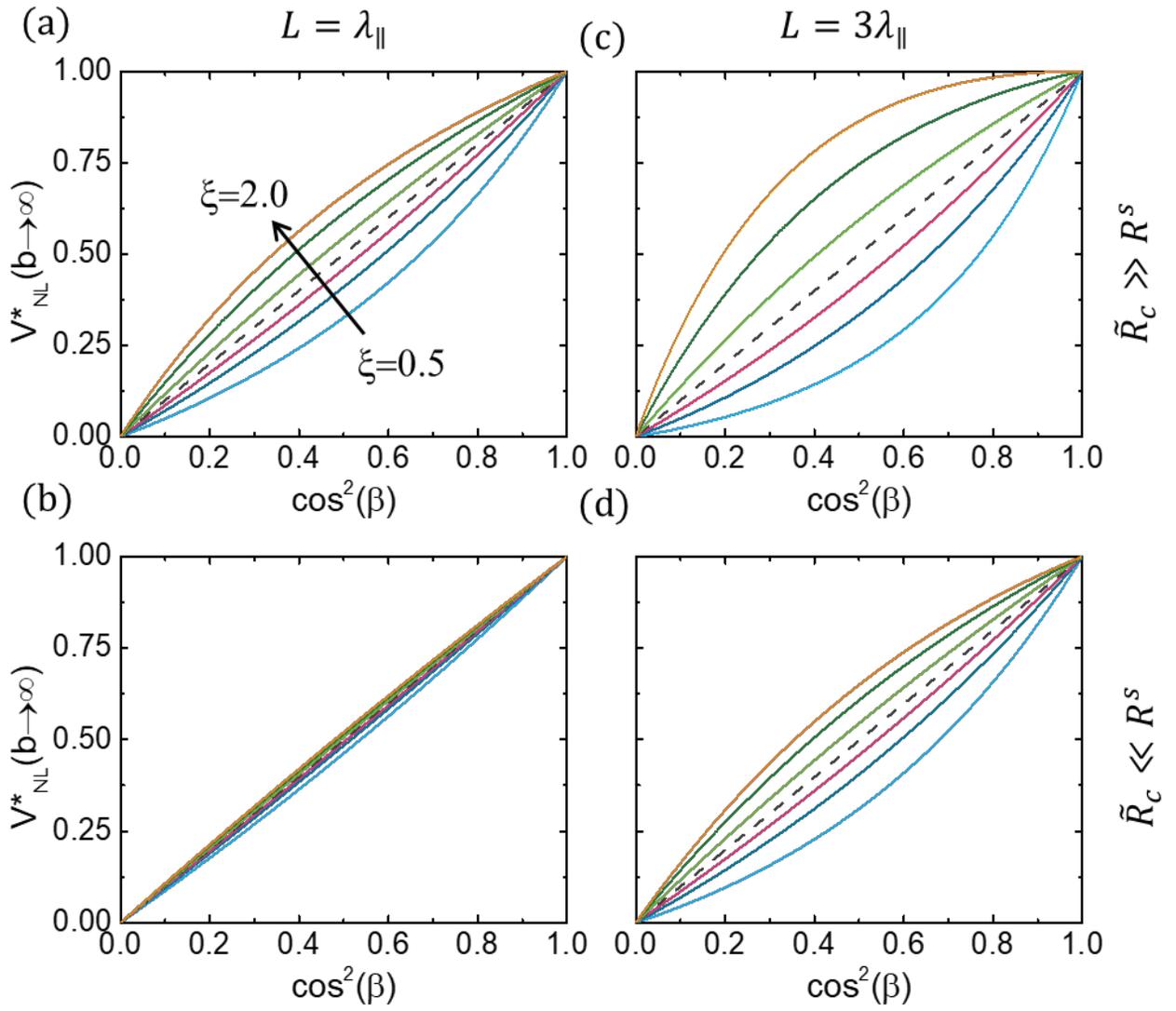

Fig. 3. Angle dependence of $V^*_{NL}(b \to \infty)$ calculated with Eq. (13), with tunneling contact (a, c) and transparent contact (b, d) resistance and different channel length. The feature of spin lifetime anisotropy is much obvious with high contact resistance and longer channel length.

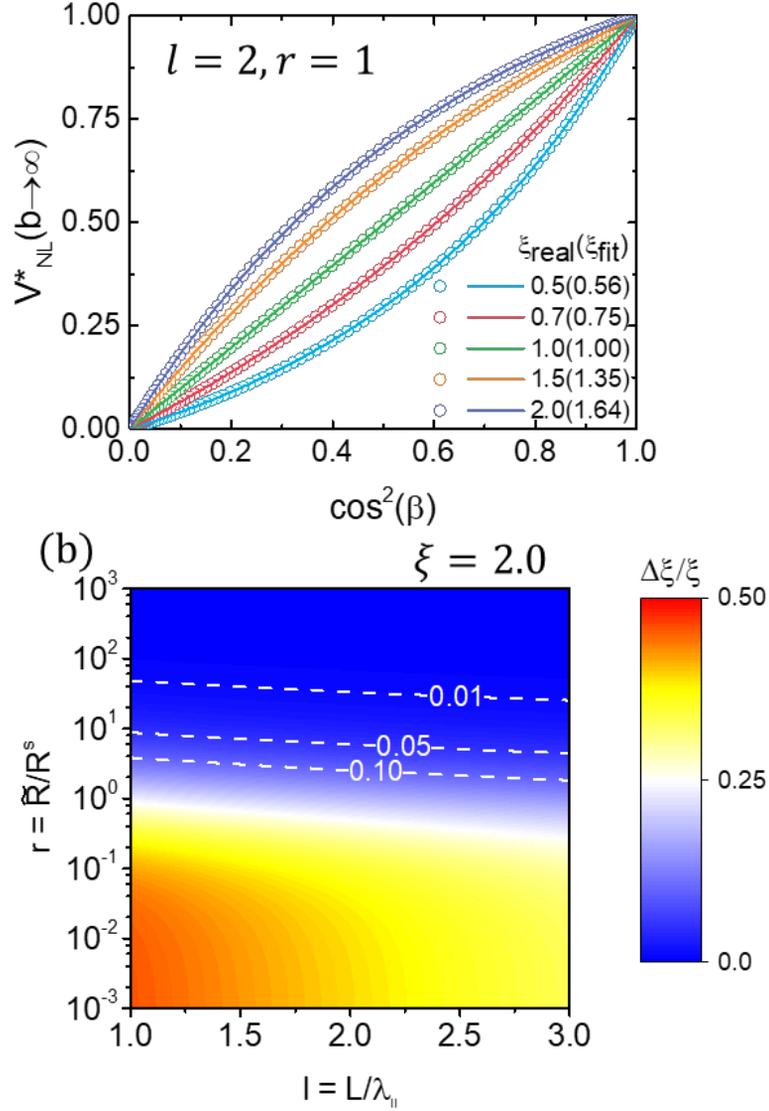

Fig. 4. (a) Fitting of the simulated oblique spin precession curves. The open circles are data from simulation, with l = 2 and r = 1. The solid lines are fittings of the simulated curve with Eq. (14), considering no contact induced spin relaxation. (b) Simulation of the fitting error $\Delta\xi/\xi$ with different contact resistance $r$ and channel length $l$. The dashed lines marked the condition when the error is 10%, 5% and 1%.

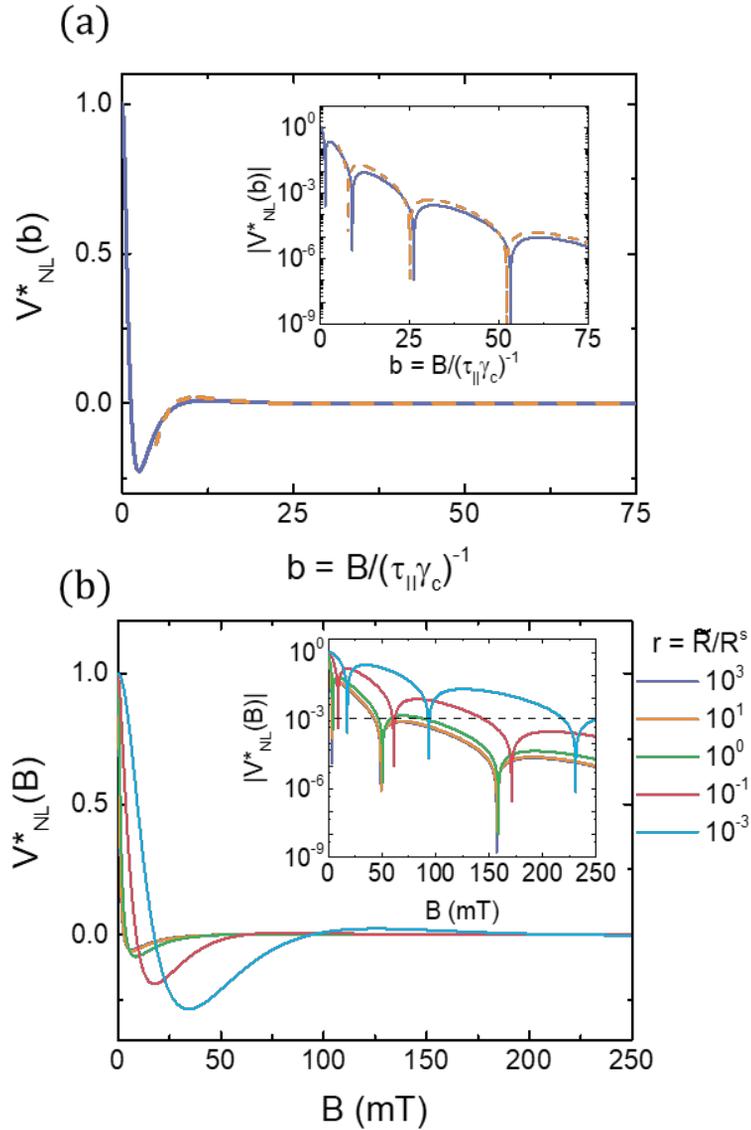

Fig. 5. (a) Simulated Hanle spin precession curve from Eq. (11) (solid) and $V_{NL}^{\perp}$ from Eq. (16) (dashed), assuming $l = 2$. Both curves are normalized with the value at $b = 0$. Because Eq. (16) is for $b \gg 1$, only the values for $b > 5$ of the dash curve is plotted. The inset shows the same curve in the semi-log scale. (b) A set of $V_{NL}^{*}$ curves simulated with different normalized contact resistance. Inset shows the same data plotted in the semi-log scale. The dashed line shows $V_{NL}^{*} = 10^{-3}$, ($\eta = 3$).